\documentclass[pdflatex,sn-mathphys-num]{sn-jnl}

\usepackage{graphicx}%
\usepackage{multirow}%
\usepackage{amsmath,amssymb,amsfonts}%
\usepackage{amsthm}%
\usepackage{mathrsfs}%
\usepackage[title]{appendix}%
\usepackage{xcolor}%
\usepackage{textcomp}%
\usepackage{manyfoot}%
\usepackage{booktabs}%
\usepackage{algorithm}%
\usepackage{algorithmicx}%
\usepackage{algpseudocode}%
\usepackage{listings}%
\usepackage{pgfplots}
\usepackage{tikz}

\theoremstyle{thmstyleone}%

\theoremstyle{thmstyletwo}%

\theoremstyle{thmstylethree}%

\begin{document}

\title[CHANRG]{Fair splits flip the leaderboard: CHANRG reveals limited generalization in RNA secondary-structure prediction}


\author[1,2,3]{\fnm{Zhiyuan} \sur{Chen}}\email{this@zyc.ai}

\author[4,5,6]{\fnm{Zhenfeng} \sur{Deng}}\email{zfevan.deng@mail.utoronto.ca}

\author[7]{\fnm{Pan} \sur{Deng}}\email{dengpan@bza.edu.cn}

\author[8]{\fnm{Yue} \sur{Liao}}\email{liaoyue.ai@gmail.com}

\author[9]{\fnm{Xiu} \sur{Su}}\email{xiusu1994@csu.edu.cn}

\author*[2,10]{\fnm{Peng} \sur{Ye}}\email{yepeng@pjlab.org.cn}

\author*[1]{\fnm{Xihui} \sur{Liu}}\email{xihuiliu@eee.hku.hk;}

\affil*[1]{\orgdiv{Department of Electrical and Computer Engineering}, \orgname{The University of Hong Kong}, \orgaddress{\street{Pok Fu Lam}, \city{Hong Kong Island}, \state{Hong Kong SAR}, \country{China}}}

\affil[2]{\orgdiv{AI4Science Center}, \orgname{Shanghai Artificial Intelligence Laboratory}, \orgaddress{\street{129 Longwen Road}, \city{Xuhui}, \postcode{200232}, \state{Shanghai}, \country{China}}}

\affil[3]{\orgdiv{Department of Pathology}, \orgname{Stanford Medicine}, \orgaddress{\street{265 Campus Drive}, \city{Stanford}, \postcode{94305}, \state{Califorina}, \country{US}}}

\affil[4]{\orgdiv{Department of Molecular Medicine},
  \orgname{The Hospital for Sick Children (SickKids), Peter Gilgan Centre for Research and Learning},
\orgaddress{\street{686 Bay Street}, \city{Toronto}, \state{Ontario}, \postcode{M5G 0A4}, \country{Canada}}}

\affil[5]{\orgdiv{Department of Molecular Genetics},
  \orgname{University of Toronto},
\orgaddress{\street{1 King's College Circle}, \city{Toronto}, \postcode{M5S 1A8}, \state{Ontario}, \country{Canada}}}

\affil[6]{\orgname{Vector Institute for Artificial Intelligence},
\orgaddress{\street{661 University Avenue}, \city{Toronto}, \postcode{M5G 1M1}, \state{Ontario}, \country{Canada}}}

\affil[7]{\orgname{Zhongguancun Academy}, \orgaddress{\street{17 Second Ring Road, Daniufang}, \city{Haidian}, \postcode{100094}, \state{Beijing}, \country{China}}}

\affil[8]{\orgdiv{School of Computing}, \orgname{National University of Singapore}, \orgaddress{\street{13 Computing Drive}, \postcode{117417}, \country{Singapore}}}

\affil[9]{\orgdiv{Big Data Institute}, \orgname{Central South University}, \orgaddress{\street{932 Lushan South Road}, \city{Changsha}, \postcode{410083}, \state{Hunan}, \country{China}}}

\affil[10]{\orgdiv{Department of Information Engineering}, \orgname{Chinese University of Hong Kong}, \orgaddress{\street{Sha Tin}, \city{New Territories}, \state{Hong Kong SAR}, \country{China}}}


\abstract{
  Accurate prediction of RNA secondary structure underpins transcriptome annotation, mechanistic analysis of non-coding RNAs, and RNA therapeutic design.
  Recent gains from deep learning and RNA foundation models are difficult to interpret because current benchmarks may overestimate generalization across RNA families.
  We present the Comprehensive Hierarchical Annotation of Non-coding RNA Groups (CHANRG), a benchmark of 170{,}083 structurally non-redundant RNAs curated from more than 10 million sequences in Rfam~15.0 using structure-aware deduplication and architecture-aware split design.
  Across 29 predictors, foundation-model methods achieved the highest held-out accuracy but lost most of that advantage out of distribution, whereas structured decoders and direct neural predictors remained markedly more robust.
  This gap persisted after controlling for sequence length and reflected both loss of structural coverage and incorrect higher-order wiring.
  Together, CHANRG and a padding-free, symmetry-aware evaluation stack provide a stricter and batch-invariant framework for developing RNA structure predictors with demonstrable out-of-distribution robustness.
}




\maketitle

\section{Introduction}\label{sec:intro}

The secondary structure of RNA, defined by its pattern of intramolecular base pairs, is a central determinant of RNA folding and function and underlies the diverse catalytic and regulatory roles of RNA molecules in biology~\cite{tinoco1999how,mortimer2014insights,kruger1982self,doudna2002although,isaacs2006rna}.
By shaping three-dimensional conformations and conformational dynamics, it also informs the design of RNA-based therapeutics and RNA-guided molecular tools~\cite{dethoff2012functional,sullenger2016from,pardi2024mrna,jinek2013rna}.
Experimental assays provide valuable structural evidence, but they remain condition-dependent and incomplete across transcripts, biological states, and structural resolutions~\cite{ding2014in,rouskin2014genome,spitale2015structural,bevilacqua2016genome}.
Computational prediction therefore complements experiments, enables transcriptome-scale annotation, and guides design in RNA therapeutics and synthetic biology~\cite{sato2023recent,wang2023uni}.

Current RNA secondary-structure predictors can be grouped operationally into three classes.
Structured decoders (SD) produce the final structure under thermodynamic, statistical, or hybrid structured optimization, as represented by EternaFold~\cite{waymentsteele2022rna}, CONTRAfold~\cite{do2006contrafold}, RNAfold~\cite{lorenz2011viennarna}, and RNAstructure~\cite{reuter2010rnastructure}.
Direct neural predictors (DL), for instance bpFold~\cite{zhu2025deep}, SPOT-RNA~\cite{singh2019rna} and UFold~\cite{fu2022ufold}, learn contact maps from sequence without a pretrained RNA language model.
Foundation-model (FM) predictors couple pretrained RNA encoders to learned structure heads~\cite{chen2022interpretable,zou2024large,penic2025rinalmo,yin2025ernie}.
Although prior work has reported improved generalization for RNA language models~\cite{penic2025rinalmo}, these results were obtained on benchmark settings that differ from the structure-aware, genome-aware, and hierarchical out-of-distribution regimes considered here, and may not fully predict transfer under stricter evaluation conditions.
Although all three classes can achieve strong held-out performance, it remains unclear whether recent gains reflect transferable structure learning or improved fitting to permissive benchmark settings~\cite{qiu2023sequence,szikszai2022deep,zhu2025deep}.

As current predictors can already achieve strong held-out performance on familiar datasets, the more pressing question is whether they generalize across families, structural regimes, and reference genomes that were not represented during model development.

Evaluation practice has not kept pace with model development~\cite{justyna2023rna,schneider2023when}. 
First, many widely used bpRNA-derived benchmark datasets were constructed from older source collections relative to recent Rfam releases~\cite{danaee2018bprna,ontiveros2025rfam}.
Second, these datasets are typically deduplicated primarily by sequence identity, so structurally similar RNAs can remain on both sides of the evaluation boundary even when primary-sequence similarity is modest~\cite{qiu2023sequence,lasher2025bprna}. 
Finally, pair-level scores can mask higher-order structural errors, including incorrect junction wiring and topological mismatches~\cite{zhao2018evaluation,mathews2019how}. 
These limitations motivate a benchmark that selects evaluation examples with genuine structural novelty relative to training, controls leakage through shared sequence or reference-genome context, and evaluates predictions hierarchically from base-pair recovery to higher-order topology.
The benchmark design in this work builds on a long lineage of curated RNA resources and annotation standards, including successive Rfam releases and curation updates~\cite{kalvari2021rfam,ontiveros2025rfam}. 
It is also informed by established RNA reference databases and analysis tools, including CRW, SRPDB, tmRDB, and VARNA~\cite{cannone2002comparative,rosenblad2003srpdb,zwieb2003tmrdb,darty2009varna}.
Recent comparative studies emphasize that robust benchmarking should test transfer beyond within-family interpolation and into new-family or low-similarity settings~\cite{qiu2023sequence,justyna2023rna,szikszai2022deep}.
They also show that pairwise overlap alone does not fully capture structural fidelity~\cite{zhao2018evaluation,mathews2019how}.

\begin{figure}[!htbp]
  \centering
  \includegraphics[width=0.9\textwidth]{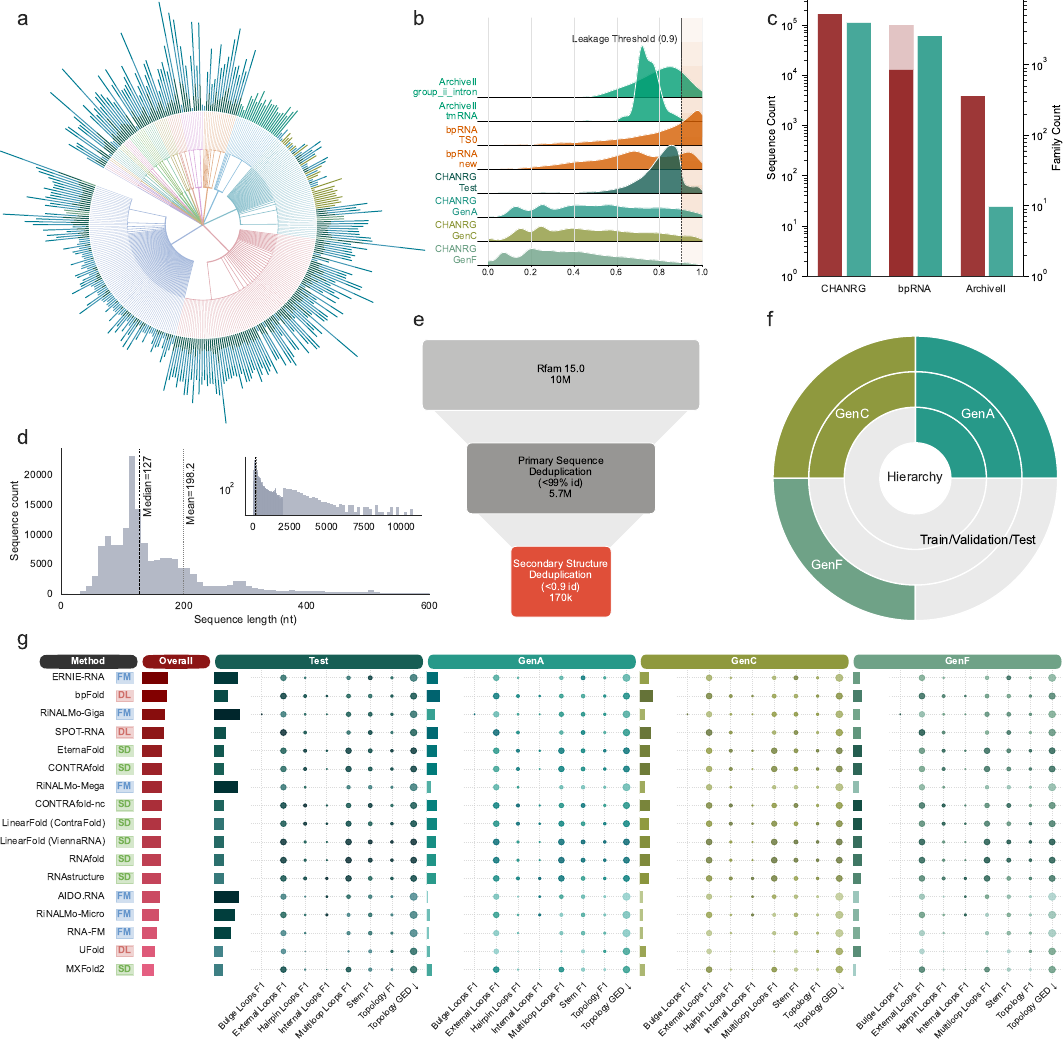}
  \caption{
    \textbf{Overview of the CHANRG benchmark.}
    \textbf{a,} Architecture--clan--family hierarchy of CHANRG with outer split occupancy across families.
    \textbf{b,} Structural similarity of CHANRG evaluation splits and widely used legacy benchmark sets relative to the CHANRG training set, showing reduced overlap under CHANRG.
    \textbf{c,} Sequence and family counts for CHANRG, bpRNA and ArchiveII-derived resources.
    \textbf{d,} Overall sequence-length distribution of CHANRG, highlighting the strongly long-tailed range of benchmarked RNAs.
    \textbf{e,} Multistage curation funnel from raw Rfam~15.0 sequences to the final structurally non-redundant benchmark.
    \textbf{f,} Biological rationale of held-out and out-of-distribution split regimes.
    \textbf{g,} First benchmark overview across representative methods, splits and structural metrics, showing that held-out leaders do not retain the same advantage across CHANRG out-of-distribution regimes.
  }
  \label{fig:main}
\end{figure}

Variable-length contact-map prediction introduces a second, less visible limitation.
Dense padded tensors waste substantial memory and computation because contact maps scale quadratically with sequence length.
When batches are padded to the longest RNA they contain, dense convolution can also make predictions depend on batch composition rather than on sequence content alone, as quantified later in this work (Fig.~\ref{fig:padding}c).
This batch-context dependence undermines reproducibility and makes benchmark-scale evaluation unnecessarily expensive (Fig.~\ref{fig:padding}c).
A rigorous benchmark for RNA secondary-structure prediction therefore requires both stronger dataset design and a compute path that respects variable-length inputs (Fig.~\ref{fig:main}b,f; Fig.~\ref{fig:padding}a--c).

Here we introduce the Comprehensive Hierarchical Annotation of Non-coding RNA Groups (CHANRG), a benchmark curated from Rfam~15.0 with structure-aware deduplication based on bpRNA-CosMoS~\cite{ontiveros2025rfam,lasher2025bprna}.
The architecture-aware split design is biologically motivated by hierarchical RNA structure classification schemes defined in RNArchitecture~\cite{boccaletto2018rnarchitecture}.
CHANRG includes held-out in-distribution Validation and Test splits together with three biologically distinct out-of-distribution regimes that probe transfer to a held-out architectural regime, clans absent from training, and genome-sparse families under limited within-family source diversity.
We benchmark 29 predictors spanning structured decoders, direct neural predictors, and foundation-model predictors under standardized preprocessing and scoring.
All foundation-model baselines were instantiated through the MultiMolecule framework~\cite{multimolecule}.
To support faithful variable-length evaluation, we also provide a padding-free, symmetry-aware reference implementation that removes padded positions from the computational graph and avoids redundant computation on symmetric contact maps.
Using CHANRG, we show that foundation-model predictors achieve the highest held-out accuracy but lose most of that advantage out of distribution, whereas structured and direct neural predictors remain markedly more robust (Fig.~\ref{fig:main_results}a--c).
Together, CHANRG and the accompanying reference implementation provide a community resource for rigorous evaluation and a practical framework for developing RNA secondary-structure predictors that generalize beyond familiar training families (Fig.~\ref{fig:main}--\ref{fig:padding}).

\section{Results}\label{sec:results}

\subsection{CHANRG: a structure-aware, leakage-controlled benchmark for RNA secondary-structure prediction}

We constructed CHANRG to address a central limitation of existing RNA secondary-structure benchmarks, as sequence-only deduplication is insufficient because RNAs with modest primary-sequence similarity can still share highly similar secondary-structure topology, and inferred structural annotations may not always be supported by evolutionary evidence~\cite{rivas2017statistical,qiu2023sequence,lasher2025bprna}.
As a result, structurally redundant examples can persist across evaluation boundaries and inflate apparent generalization.
CHANRG therefore explicitly removes both sequence and structural redundancy while combining held-out in-distribution splits with biologically distinct out-of-distribution regimes.
Evaluation uses held-out in-distribution Validation and Test splits together with three out-of-distribution regimes that probe transfer to a held-out architectural regime (GenA), clans absent from training (GenC), and genome-sparse families under limited within-family source diversity (GenF).
Performance is assessed using a multiscale metric ladder spanning base-pair $F_1$ for local contact recovery, stem $F_1$ for helix-level recovery, topology $F_1$ for higher-order structural organization, and topology GED, a lower-is-better structural edit distance~\cite{sanfeliu1983distance}.
An overview of benchmark construction, split design, dataset properties, metric hierarchy, and the first benchmark summary appears in Fig.~\ref{fig:main}.

Starting from Rfam release~15.0, we applied a multi-stage curation pipeline to 10{,}025{,}911 sequences drawn from 4{,}178 source families~\cite{ontiveros2025rfam}.
After integrity screening, 10{,}025{,}740 sequences remained, followed by 5{,}670{,}054 after sequence-level deduplication and 170{,}083 after structure-aware deduplication.
This pipeline comprised integrity screening, high-stringency sequence-level deduplication, and structure-aware deduplication based on bpRNA-CosMoS similarity scores~\cite{lasher2025bprna}.
Thus, even after stringent sequence-level filtering, structure-aware pruning removed an additional 33-fold of residual redundancy, indicating that many non-identical RNAs still shared highly similar secondary structures.
Figure~\ref{fig:main}e summarizes this curation funnel, whereas Fig.~\ref{fig:main}c situates CHANRG relative to widely used legacy resources in both sequence count and family count.

To test whether structural leakage remained after curation, we compared structural-similarity distributions of CHANRG evaluation sets against the training set with those of commonly used legacy benchmark sets, including bpRNA-derived and ArchiveII family-fold settings~\cite{danaee2018bprna,sloma2016exact,singh2019rna,qiu2023sequence}.
These distributions showed that CHANRG evaluation splits are less structurally coupled to training than these legacy resources (Fig.~\ref{fig:main}b).
This distinction is important because sequence-level filtering alone can still leave highly similar folds on both sides of an evaluation boundary~\cite{qiu2023sequence,lasher2025bprna}.
Structure-aware curation therefore changes not only the size of the benchmark, but also the effective novelty of the examples used to assess generalization.

Split design combines the hierarchical organization of non-coding RNAs with a reference-genome-aware rule that separates development from evaluation within families~\cite{kalvari2021rfam,ontiveros2025rfam}.
Figure~\ref{fig:main}a visualizes the architecture--clan--family hierarchy of CHANRG, and Fig.~\ref{fig:main}f summarizes the biological rationale of the held-out and out-of-distribution splits.
GenA contains sequences annotated as ``complex unclassified'' and therefore probes transfer to a held-out architectural regime.
GenC contains sequences from clans absent from training and therefore probes broader evolutionary distance beyond the training clan hierarchy.
For the remaining families, Validation and Test are constructed so that no two sequences from the same reference genome appear together within a family.
Families that cannot provide sufficient genome diversity for this split are assigned to GenF, yielding a distinct family-level stress test under sparse phylogenetic coverage.
Sequences not assigned to Validation, Test, or one of the three OOD regimes are retained for training.

The final dataset comprises 123{,}223 Train sequences, 14{,}070 Validation sequences, 14{,}070 Test sequences, 12{,}499 GenA sequences, 4{,}424 GenC sequences, and 1{,}797 GenF sequences.
Sequence lengths are strongly long-tailed, with split-specific medians of 128 nt in Test, 211 nt in GenA, 93 nt in GenC, and 89 nt in GenF (Fig.~\ref{fig:main}d).
Pseudoknot prevalence is low overall but heterogeneous across evaluation regimes, reaching 2.8\% in Test and 2.7\% in GenA, compared with 0.2\% in GenC and 0.3\% in GenF.
Unlike legacy benchmarks built primarily around sequence-level curation, CHANRG combines structural deduplication with biologically distinct OOD splits~\cite{danaee2018bprna,qiu2023sequence,lasher2025bprna}.
Together, this scale, structural diversity, and split design create a stringent test bed for RNA secondary-structure generalization, and the first benchmark overview in Fig.~\ref{fig:main}g previews how these design choices reshape model comparisons.

\begin{figure}[!htbp]
  \centering
  \includegraphics[width=0.9\textwidth]{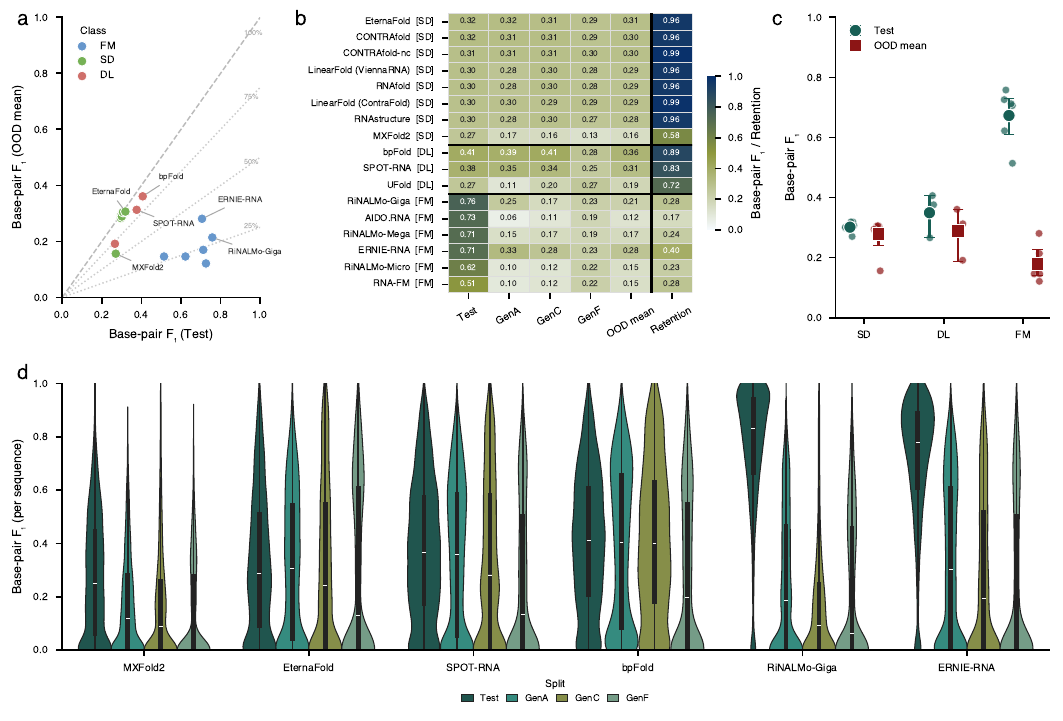}
  \caption{
    \textbf{Standard held-out leaderboards overestimate generalization.}
    \textbf{a,} Test versus OOD$_{\mathrm{mean}}$ base-pair $F_1$ for the canonical 17-model cohort. Each point is one model, colored by predictor class (FM, SD, or DL). Dashed guide lines indicate equal performance and 75\%, 50\%, and 25\% retention relative to Test.
    \textbf{b,} Heatmap of base-pair $F_1$ across Test, GenA, GenC, and GenF, together with OOD$_{\mathrm{mean}}$ and retention (OOD$_{\mathrm{mean}}$/Test), for the canonical cohort. Held-out leaders lose their advantage across CHANRG OOD regimes.
    \textbf{c,} Class-level base-pair $F_1$ on Test and OOD$_{\mathrm{mean}}$. Points denote individual models and error bars denote 95\% bootstrap confidence intervals over per-model means.
    \textbf{d,} Per-sequence base-pair $F_1$ distributions across Test, GenA, GenC, and GenF for six representative methods: MXFold2 and EternaFold (SD), SPOT-RNA and BPfold (DL), and RiNALMo-Giga and ERNIE-RNA (FM). OOD$_{\mathrm{mean}}$ denotes the unweighted mean across GenA, GenC, and GenF.
  }
  \label{fig:main_results}
\end{figure}

\subsection{Standard held-out leaderboards overestimate generalization}

We next benchmarked a canonical 17-model cohort comprising 8 structured decoders, 3 direct neural predictors, and 6 foundation-model predictors.
Structured decoders produce the final structure under explicit folding constraints or structured optimization, direct neural predictors infer structure directly from sequence without a pretrained RNA language model, and foundation-model predictors combine a pretrained RNA encoder with a learned structure head.
To avoid overweighting multiple structure heads attached to the same pretrained backbone, class-level summaries use one U-Net head per foundation-model backbone or scale~\cite{ronneberger2015unet}, whereas full results for all 29 evaluated models are reported in Extended Data.
To summarize out-of-distribution behavior, we define $\mathrm{OOD}_{\mathrm{mean}}$ as the unweighted mean across GenA, GenC, and GenF.

Held-out leaderboards and out-of-distribution evaluation favor different predictor classes.
Foundation-model predictors achieved the highest mean Test base-pair $F_1$, reaching $0.6731$ across the canonical cohort, whereas direct neural predictors reached $0.3495$ and structured decoders reached $0.3015$ (Fig.~\ref{fig:main_results}c).
However, foundation-model performance fell to $0.1796$ on $\mathrm{OOD}_{\mathrm{mean}}$, corresponding to an absolute loss of $0.4935$ and a retention of only $26.7\%$ relative to Test.
By contrast, direct neural predictors retained $82.5\%$ of their Test performance, decreasing from $0.3495$ to $0.2883$, whereas structured decoders retained $92.3\%$, decreasing only from $0.3015$ to $0.2784$.
Conventional held-out evaluation therefore substantially overestimates the robustness of foundation-model predictors (Fig.~\ref{fig:main_results}a--c).
This observation contrasts with prior reports of improved generalization in RNA language models~\cite{penic2025rinalmo}, suggesting that benchmark design critically affects the measured transfer behavior.

Model-level comparisons make this inversion visually explicit.
In the canonical 17-model cohort, foundation-model predictors cluster toward high Test accuracy but weak $\mathrm{OOD}_{\mathrm{mean}}$ performance, whereas structured decoders and direct neural predictors lie closer to the retention diagonal (Fig.~\ref{fig:main_results}a).
The corresponding heatmap shows that held-out leaders lose their advantage across GenA, GenC, and GenF rather than on only one OOD split (Fig.~\ref{fig:main_results}b).
Among foundation-model predictors, RiNALMo-giga-U-Net~\cite{penic2025rinalmo} achieved the highest Test base-pair $F_1$ of $0.7579$, but its performance fell to $0.2509$ on GenA, $0.1651$ on GenC, and $0.2260$ on GenF, for an overall $\mathrm{OOD}_{\mathrm{mean}}$ of $0.2140$.
Within the main-text foundation-model cohort, ERNIE-RNA-U-Net~\cite{yin2025ernie} provided the strongest out-of-distribution performance, reaching $\mathrm{OOD}_{\mathrm{mean}} = 0.2807$.
Among structured decoders, EternaFold~\cite{waymentsteele2022rna} achieved the strongest aggregate OOD performance, with a Test base-pair $F_1$ of $0.3189$ and an $\mathrm{OOD}_{\mathrm{mean}}$ of $0.3064$, whereas RNAfold~\cite{lorenz2011viennarna} remained comparably stable at $0.3013$ on Test and $0.2886$ on $\mathrm{OOD}_{\mathrm{mean}}$.
Among direct neural predictors, BPfold~\cite{zhu2025deep} achieved the strongest overall OOD performance, with a Test score of $0.4065$ and an $\mathrm{OOD}_{\mathrm{mean}}$ of $0.3608$.
The highest-scoring model on the conventional Test split therefore did not belong to the class that generalized best across architectural, clan-level, and family-level shift.

Bootstrap resampling over per-model means yielded the same qualitative inversion, indicating that the class-level contrast was not driven by a single model variant.
On Test, the 95\% confidence interval for the foundation-model class $[0.6004, 0.7300]$ did not overlap the structured-decoder interval $[0.2910, 0.3106]$ (Fig.~\ref{fig:main_results}c).
On $\mathrm{OOD}_{\mathrm{mean}}$, this relationship reversed, with the foundation-model interval $[0.1417, 0.2247]$ lying entirely below the structured-decoder interval $[0.2418, 0.3004]$.
The same inversion held at higher structural abstraction.
For topology $F_1$, which measures recovery of stems, loops, and their connections, foundation-model predictors dominated on Test $[0.3517, 0.4870]$ but not on $\mathrm{OOD}_{\mathrm{mean}}$, where their interval $[0.0550, 0.0935]$ fell below the structured-decoder interval $[0.0982, 0.1208]$ (Fig.~\ref{fig:main_results}d).
For topology GED, a lower-is-better structural edit distance on the loop--helix graph, foundation-model predictors were best on Test $[0.3212, 0.4598]$ but worst on $\mathrm{OOD}_{\mathrm{mean}}$, where their interval $[0.6873, 0.7598]$ lay well above the structured-decoder interval $[0.5111, 0.5462]$.
Thus, the apparent superiority of foundation-model predictors under held-out evaluation inverts systematically under CHANRG's out-of-distribution regimes.

Held-out leaderboard rank is also a poor proxy for out-of-distribution robustness within the foundation-model class.
Among foundation-model predictors, the Spearman correlation between Test rank and $\mathrm{OOD}_{\mathrm{mean}}$ rank was weak ($\rho = 0.200$, $P = 0.704$).
By contrast, structured decoders preserved their ranking much more consistently across held-out and OOD evaluation ($\rho = 0.905$, $P = 2.0\times10^{-3}$).
Because the direct-neural class contains only three methods, we do not overinterpret within-class rank stability for that group.
These results indicate that leaderboard position on standard held-out sets is a poor proxy for cross-regime robustness among foundation-model predictors, whereas structured decoders remain substantially more stable (Fig.~\ref{fig:main_results}a--c).

\begin{figure}[!htbp]
  \centering
  \includegraphics[width=0.9\textwidth]{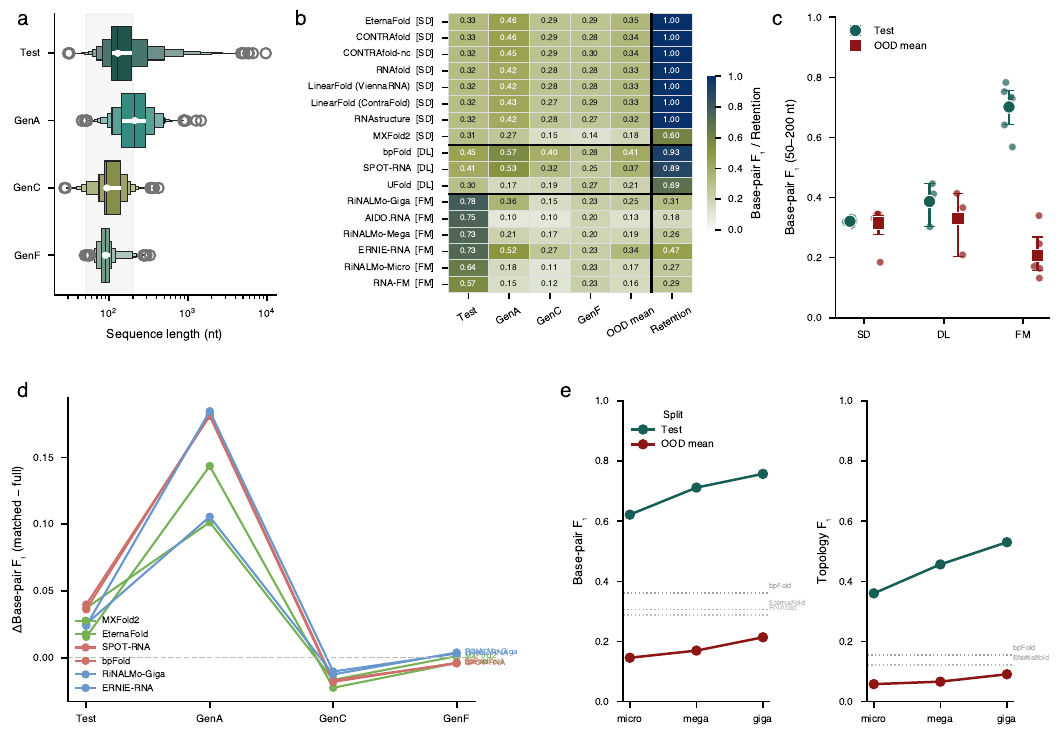}
  \caption{
    \textbf{The CHANRG generalization gap is not explained by sequence length or model scale.}
    \textbf{a,} Split-specific sequence-length distributions (log-scale x-axis) show that GenA is longer than Test, whereas GenC and GenF are shorter, ruling out a simple ``OOD = longer RNAs'' explanation.
    \textbf{b,} Base-pair $F_1$ heatmap for the canonical 17-model cohort after restricting all splits to RNAs between 50 and 200 nt, with OOD$_{\mathrm{mean}}$ and retention columns.
    \textbf{c,} Class-level base-pair $F_1$ on the length-matched subset, showing that the held-out/OOD inversion persists after controlling the evaluated length range. Points denote individual models and error bars denote 95\% bootstrap confidence intervals over per-model means.
    \textbf{d,} Change in base-pair $F_1$ after length matching ($F_1^{\mathrm{matched}} - F_1^{\mathrm{full}}$) for six representative methods. Length control improves several structured and direct neural baselines on GenA but does not rescue foundation-model transfer on GenC or GenF.
    \textbf{e,} Scaling within the RiNALMo-U-Net family improves Test performance much more than OOD performance for both base-pair $F_1$ and topology $F_1$. Horizontal reference lines indicate BPfold, EternaFold, and RNAfold on the corresponding metric. OOD$_{\mathrm{mean}}$ denotes the unweighted mean across GenA, GenC, and GenF.
  }
  \label{fig:length_scale}
\end{figure}

\subsection{The generalization gap is not explained by sequence length}

A natural concern is that the OOD deficit simply reflects differences in sequence length across splits.
That explanation is incomplete.
GenA is longer than Test, with median lengths of 211 and 128 nt, respectively, but GenC and GenF are both shorter than Test, with median lengths of 93 and 89 nt (Fig.~\ref{fig:length_scale}a).
If length were the dominant driver of OOD failure, performance should recover on the shorter OOD regimes, yet large deficits persist there.
Split composition therefore suggests that length contributes to difficulty, but cannot by itself explain the overall generalization pattern.

We next repeated the benchmark after restricting all splits to RNAs between 50 and 200 nt.
The main class-level result persisted under this control (Fig.~\ref{fig:length_scale}b).
Within the canonical cohort, foundation-model predictors achieved a mean Test base-pair $F_1$ of $0.7016$ but only $0.2074$ on $\mathrm{OOD}_{\mathrm{mean}}$ under length matching.
By contrast, structured decoders achieved $0.3212$ on Test and $0.3156$ on $\mathrm{OOD}_{\mathrm{mean}}$, whereas direct neural predictors achieved $0.3870$ on Test and $0.3299$ on $\mathrm{OOD}_{\mathrm{mean}}$.
Thus, controlling the evaluated length range does not rescue foundation-model performance out of distribution.

Model-level comparisons yielded the same conclusion.
For RiNALMo-giga-U-Net, base-pair $F_1$ increased from $0.7579$ to $0.7832$ on Test and from $0.2509$ to $0.3565$ on GenA after length matching, but remained very low on GenC at $0.1525$ and on GenF at $0.2298$.
By contrast, several structured and direct neural baselines improved substantially on GenA under the same control, including BPfold from $0.3876$ to $0.5687$, RNAfold from $0.2802$ to $0.4193$, and EternaFold from $0.3171$ to $0.4608$ (Fig.~\ref{fig:length_scale}c).
Length matching therefore reveals an informative class asymmetry: long sequences partly penalize some structured and direct neural methods on GenA, whereas foundation-model failure persists on the shorter GenC and GenF regimes.

These results argue against a simple geometric explanation for the held-out versus OOD inversion.
The decisive observation is GenC: although its sequences are shorter than those in Test, foundation-model predictors remain much worse there than on held-out evaluation.
The generalization deficit revealed by CHANRG is therefore better explained by structural and evolutionary shift than by sequence length alone.
We next asked whether the same benchmark result might instead reflect insufficient foundation-model scale.

\subsection{Scaling foundation models improves held-out accuracy more than out-of-distribution robustness}

We next examined whether increasing foundation-model capacity is sufficient to close the generalization gap revealed by CHANRG.
To vary scale while holding the backbone family and prediction head fixed, we compared RiNALMo-micro-U-Net, RiNALMo-mega-U-Net, and RiNALMo-giga-U-Net (Fig.~\ref{fig:length_scale}d).
This comparison isolates scale from head choice and therefore provides a direct test of whether larger foundation models solve the benchmark's central OOD problem.

Scaling substantially improved held-out base-pair accuracy, with Test $F_1$ increasing from $0.6222$ in RiNALMo-micro-U-Net to $0.7122$ in RiNALMo-mega-U-Net and $0.7579$ in RiNALMo-giga-U-Net.
The corresponding gains on $\mathrm{OOD}_{\mathrm{mean}}$ were much smaller, increasing only from $0.1460$ to $0.1697$ and $0.2140$, respectively.
From micro to giga, scale therefore improved Test base-pair $F_1$ by $0.1357$, but improved $\mathrm{OOD}_{\mathrm{mean}}$ by only $0.0680$.
The same asymmetry was visible across the individual OOD regimes: scaling improved GenA substantially, but produced little improvement on GenC and only modest improvement on GenF.

Topology-aware evaluation showed the same pattern more strongly.
Test topology $F_1$ increased from $0.3602$ in RiNALMo-micro-U-Net to $0.4566$ in RiNALMo-mega-U-Net and $0.5304$ in RiNALMo-giga-U-Net.
Over the same range, $\mathrm{OOD}_{\mathrm{mean}}$ topology $F_1$ increased only from $0.0582$ to $0.0665$ and $0.0911$.
Topology GED likewise improved more on Test than out of distribution, decreasing from $0.4617$ to $0.3608$ and $0.2722$ on Test, but only from $0.7594$ to $0.7248$ and $0.6570$ on $\mathrm{OOD}_{\mathrm{mean}}$.
Thus, increasing foundation-model scale improves held-out structural reconstruction much more than transferable higher-order organization.

Even the largest RiNALMo-U-Net model remained less robust out of distribution than several lower-scoring baselines.
RiNALMo-giga-U-Net achieved the highest Test base-pair $F_1$ in the benchmark at $0.7579$, yet its $\mathrm{OOD}_{\mathrm{mean}}$ of $0.2140$ remained well below BPfold ($0.3608$), EternaFold ($0.3064$), and RNAfold ($0.2886$).
The same ranking held at higher structural abstraction, where RiNALMo-giga-U-Net reached $\mathrm{OOD}_{\mathrm{mean}}$ topology $F_1 = 0.0911$, compared with $0.1552$ for BPfold, $0.1227$ for EternaFold, and $0.1207$ for RNAfold, and had a worse $\mathrm{OOD}_{\mathrm{mean}}$ topology GED ($0.6570$) than all three baselines.
Increasing capacity therefore does not resolve the central benchmark result.
Within the current foundation-model pipeline, increasing scale improves held-out fitting much more than out-of-distribution robustness.
Because neither sequence length nor model scale is sufficient to explain the observed OOD gap, we next examined how these failures appear across the structural abstraction ladder.

\begin{figure}[!htbp]
  \centering
  \includegraphics[width=0.9\textwidth]{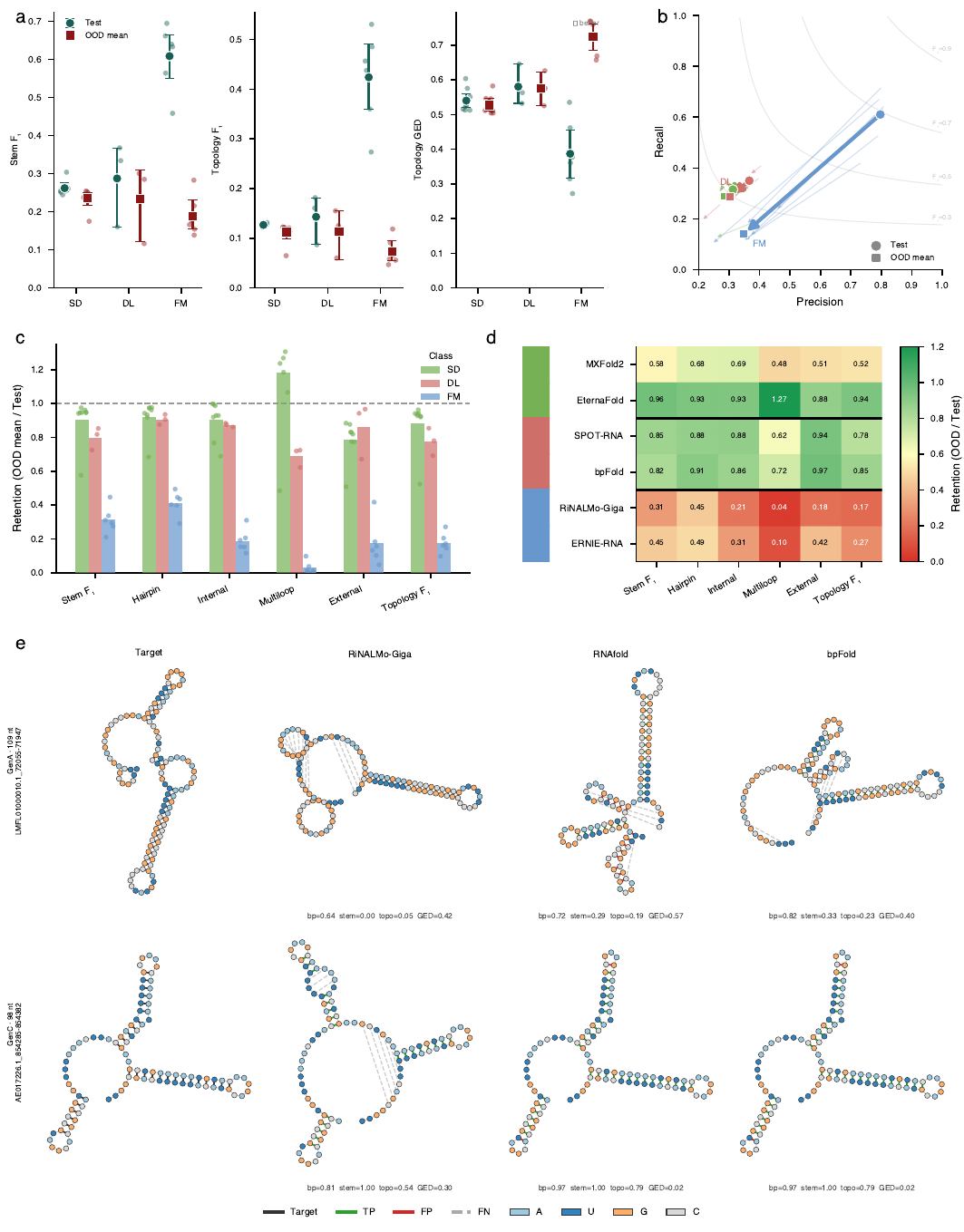}
  \caption{
    \textbf{Hierarchical evaluation reveals coverage failure and helix miswiring.}
    \textbf{a,} Class-level stem $F_1$, topology $F_1$, and topology GED on Test and OOD$_{\mathrm{mean}}$. Points denote individual models and error bars denote 95\% bootstrap confidence intervals over per-model means. Lower topology GED indicates better agreement.
    \textbf{b,} Precision--recall shift from Test to OOD$_{\mathrm{mean}}$ for structured decoders (SD), direct neural predictors (DL), and foundation-model predictors (FM). Gray contours indicate iso-$F_1$ values.
    \textbf{c,} Retention relative to Test for stem $F_1$, hairpin-loop $F_1$, internal-loop $F_1$, multiloop $F_1$, external-loop $F_1$, and topology $F_1$, stratified by method class. Points denote individual models.
    \textbf{d,} Motif-specific retention (OOD$_{\mathrm{mean}}$/Test) for six representative methods: MXFold2 and EternaFold (SD), SPOT-RNA and BPfold (DL), and RiNALMo-Giga and ERNIE-RNA (FM).
    \textbf{e,} Two pre-specified case studies localizing OOD errors to global loop--helix organization. Left, Case 1 (GenA, 109 nt; RF01527; \texttt{LMFL01000010.1\_72055-71947}) shows moderate contact recovery but incorrect global architecture for RiNALMo-Giga. Right, Case 2 (GenC, 98 nt; RF03162; \texttt{AE017226.1\_854285-854382}) shows perfect helix recovery but failed multiloop wiring. Target pairs, true positives (TP), false positives (FP), and false negatives (FN) are shown for each prediction. OOD$_{\mathrm{mean}}$ denotes the unweighted mean across GenA, GenC, and GenF.
  }
  \label{fig:failure}
\end{figure}

\subsection{Hierarchical evaluation reveals coverage failure and helix miswiring}

Base-pair $F_1$ established the existence of a generalization gap, but it understated its structural depth.
To localize this gap, we moved from pair-level accuracy to a multiscale set of structural diagnostics spanning stem $F_1$, topology $F_1$, topology GED, precision and recall, and motif-level performance (Fig.~\ref{fig:failure}a).
Here, stem $F_1$ asks whether the correct helical segments are recovered, topology $F_1$ asks whether stems, loops, and their connections are assembled into the correct higher-order organization, and topology GED measures how far the predicted loop--helix graph lies from the reference, with lower values indicating better agreement.
A model can therefore achieve high stem $F_1$ while still failing topologically if it recovers the right helices but connects them incorrectly.
These diagnostics revealed two distinct failure modes among foundation-model predictors, namely coverage failure, in which true interactions are omitted, and wiring failure, in which recovered helices are assembled into incorrect global architectures.

The strongest foundation-model predictor illustrates the scale of this collapse.
For RiNALMo-giga-U-Net, pair exact match decreased from $0.0904$ on Test to $0.0009$ on $\mathrm{OOD}_{\mathrm{mean}}$.
Over the same comparison, stem $F_1$ decreased from $0.6947$ to $0.2157$, topology $F_1$ decreased from $0.5304$ to $0.0910$, and topology GED worsened from $0.2722$ to $0.6570$.
At the class level, foundation-model predictors dropped from mean topology $F_1 = 0.4240$ on Test to $0.0729$ on $\mathrm{OOD}_{\mathrm{mean}}$, while topology GED increased from $0.3869$ to $0.7245$ (Fig.~\ref{fig:failure}b).
By contrast, structured decoders changed much less, from mean topology $F_1 = 0.1269$ on Test to $0.1125$ on $\mathrm{OOD}_{\mathrm{mean}}$, and from mean topology GED $= 0.5401$ to $0.5270$, whereas direct neural predictors changed from $0.1431$ to $0.1136$ in topology $F_1$ and from $0.5803$ to $0.5756$ in topology GED.
Thus, the OOD deficit of foundation-model predictors is not only a pair-recovery problem, but a larger collapse in higher-order structural organization.

The gap between stem $F_1$ and topology $F_1$ further localizes where this collapse occurs.
Across foundation-model predictors, mean stem $F_1$ fell from $0.6087$ on Test to $0.1890$ on $\mathrm{OOD}_{\mathrm{mean}}$, corresponding to $31.1\%$ retention.
Over the same comparison, topology $F_1$ fell from $0.4240$ to $0.0729$, corresponding to only $17.2\%$ retention.
By contrast, structured decoders retained $89.8\%$ of stem $F_1$ and $88.6\%$ of topology $F_1$, while direct neural predictors retained $81.4\%$ and $79.4\%$, respectively.
Foundation-model predictors therefore preserve some helical signal out of distribution, but fail much more severely when those helices must be assembled into the correct global wiring diagram.

Decomposing base-pair $F_1$ into precision and recall revealed the first failure mode, coverage failure.
Across foundation-model predictors, mean precision and recall shifted from $0.7968$ and $0.6108$ on Test to $0.3477$ and $0.1403$ on $\mathrm{OOD}_{\mathrm{mean}}$, moving the class toward a markedly high-precision, low-recall regime (Fig.~\ref{fig:failure}c).
The corresponding precision-to-recall ratio increased from $1.30$ to $2.48$, indicating that OOD predictions became increasingly conservative and omitted many true interactions.
By contrast, structured decoders remained comparatively balanced, with mean precision and recall changing only from $0.3121$ and $0.3149$ on Test to $0.2840$ and $0.2893$ on $\mathrm{OOD}_{\mathrm{mean}}$.
Direct neural predictors also remained far more balanced than foundation-model predictors, shifting from $0.3660$ and $0.3501$ on Test to $0.3019$ and $0.2877$ on $\mathrm{OOD}_{\mathrm{mean}}$.
Coverage failure is therefore a prominent characteristic of foundation-model predictors rather than a generic feature of all predictor classes.

Motif-level retention quantified the second failure mode, wiring failure, at scale.
Relative to Test, foundation-model predictors retained $31.1\%$ of stem $F_1$, $31.3\%$ of hairpin-loop $F_1$, $10.1\%$ of internal-loop $F_1$, $23.5\%$ of multiloop $F_1$, $48.8\%$ of external-loop $F_1$, and $17.2\%$ of topology $F_1$ on $\mathrm{OOD}_{\mathrm{mean}}$ (Fig.~\ref{fig:failure}d).
The weakest retention therefore occurred precisely in junction-dependent and higher-order motifs, especially internal loops and multiloops.
By contrast, structured decoders retained $89.8\%$ of stem $F_1$, $131.3\%$ of hairpin-loop $F_1$, $89.8\%$ of internal-loop $F_1$, $95.2\%$ of multiloop $F_1$, $92.8\%$ of external-loop $F_1$, and $88.6\%$ of topology $F_1$.
The apparent gain in hairpin-loop retention for structured decoders likely reflects relative enrichment of simpler hairpin-dominated structures in the OOD regimes rather than a literal increase in raw predictive capacity.
Overall, these motif-level results show that the OOD gap in foundation-model predictors is concentrated in structural contexts that depend on correct higher-order assembly rather than local helix detection alone.

Two pre-specified case studies localized these errors to global loop--helix organization (Fig.~\ref{fig:failure}e).
In a $109$-nt GenA RNA from RF01527 (\texttt{LMFL01000010.1\_72055-71947}), RiNALMo-giga-U-Net achieved base-pair $F_1 = 0.638$ but topology $F_1 = 0.045$, stem $F_1 = 0.000$, and topology GED $= 0.417$.
On the same RNA, RNAfold achieved base-pair $F_1 = 0.719$, topology $F_1 = 0.185$, and stem $F_1 = 0.286$, whereas BPfold achieved base-pair $F_1 = 0.821$, topology $F_1 = 0.227$, and stem $F_1 = 0.333$.
Thus, the foundation-model recovered many individual contacts, but assembled them into a fundamentally different architecture.
In a $98$-nt GenC RNA from RF03162 (\texttt{AE017226.1\_854285-854382}), RiNALMo-giga-U-Net correctly recovered all helices ($\mathrm{stem}\;F_1 = 1.000$) and still achieved base-pair $F_1 = 0.808$, but failed completely on the multiloop junction ($\mathrm{multiloop}\;F_1 = 0.000$), yielding topology $F_1 = 0.542$ and topology GED $= 0.302$.
By contrast, RNAfold and BPfold both achieved base-pair $F_1 = 0.969$, stem $F_1 = 1.000$, topology $F_1 = 0.792$, and topology GED $= 0.020$, with substantial multiloop recovery in both cases.
These examples show that many OOD errors do not arise from complete structural ignorance.
Instead, foundation-model predictors often recover local helices while failing to place them into the correct global arrangement.
Together, these analyses show that OOD failure is a coupled loss of structural coverage and structural wiring fidelity.

\begin{figure}[!htbp]
  \centering
  \includegraphics[width=0.9\textwidth]{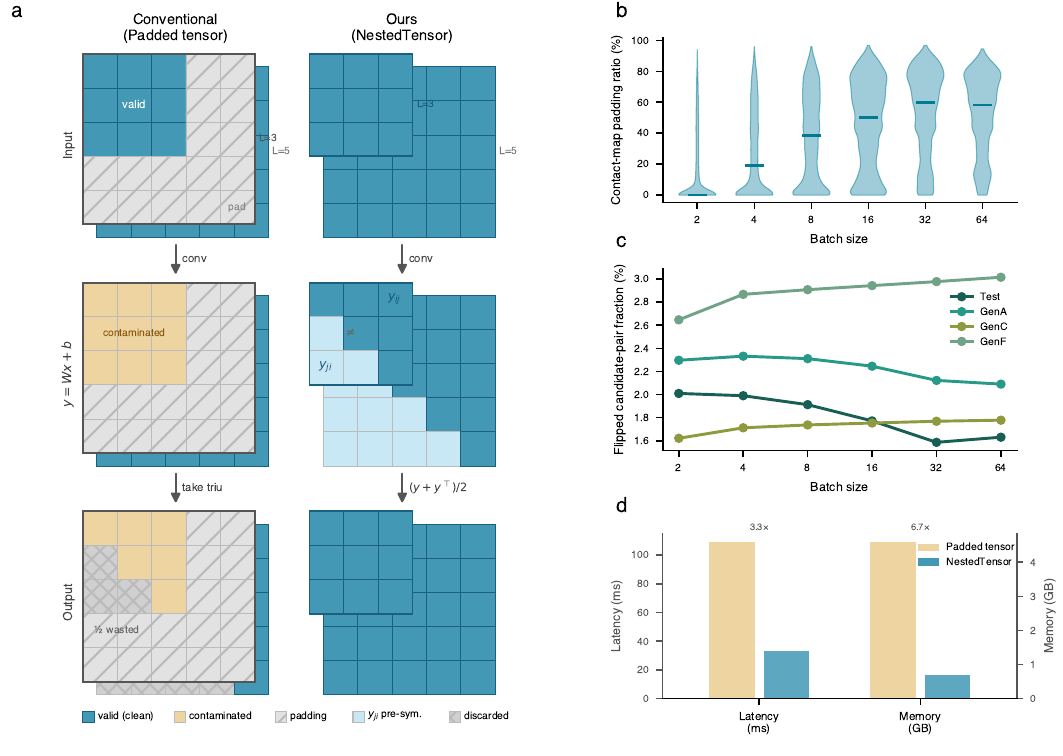}
  \caption{
    \textbf{Padding-free, symmetry-aware computation enables faithful CHANRG-scale evaluation.}
    \textbf{a,} Comparison of conventional padded-tensor execution and the NestedTensor reference implementation.
    In conventional dense execution, shorter RNAs are embedded in a larger $L_{\max}\times L_{\max}$ contact map, so convolution mixes valid and padded regions and symmetry enforcement by taking the upper triangle discards half of the dense output.
    In the NestedTensor path, padded positions are excluded from the computation graph and symmetry is enforced at the output level by averaging the predicted contact map with its transpose, $o=(y+y^{\top})/2$.
    \textbf{b,} Contact-map padding ratio across real evaluation-set batch contexts from Test, GenA, GenC, and GenF, stratified by batch size.
    \textbf{c,} Flipped candidate-pair fraction under dense padded inference across batch sizes for Test, GenA, GenC, and GenF, showing that batch-context dependence persists even at batch size 2.
    \textbf{d,} Controlled systems benchmark on synthetic batches spanning representative sequence lengths.
    Relative to dense padded tensors, the NestedTensor implementation reduces inference latency by 3.3-fold and allocated GPU memory by 6.7-fold.
    Panels \textbf{b} and \textbf{c} use real evaluation-set contexts, whereas panel \textbf{d} uses synthetic batches to isolate systems cost.
  }
  \label{fig:padding}
\end{figure}

\subsection{Padding-free, symmetry-aware computation enables faithful CHANRG-scale evaluation}

CHANRG spans RNAs from 19 to 10{,}799 nt.
Because contact maps scale quadratically with sequence length, this broad range produces extreme variation in contact-map area and makes dense padding especially costly in heterogeneous batches.
Under dense tensor execution, each sequence is padded to the longest RNA in the batch, so the wasted contact-map area for sequence $i$ is $L_{\max}^2 - L_i^2$.
Padding therefore consumes a substantial fraction of the compute budget and creates a setting in which predictions for the same RNA can depend on which longer RNAs happen to share its batch (Fig.~\ref{fig:padding}a).

We first quantified the padding burden across CHANRG evaluation.
Across 185{,}472 sequence--batch contexts derived from 32{,}416 evaluable sequences in Test, GenA, GenC, and GenF, the mean padding ratio was 0.380, 40.1\% of contexts contained more than 50\% padding, and 14.5\% contained more than 75\% padding (Fig.~\ref{fig:padding}b).
Mean padding increased from 0.152 at batch size 2 to 0.534 at batch size 32, and on the Test split reached 0.647 at batch size 32.
Dense evaluation therefore spends a large fraction of its compute budget on padded contact-map area under realistic CHANRG batching.

We next asked whether dense padding changes model predictions and whether this effect can be mitigated simply by reducing batch size.
Using an ERNIE-RNA model finetuned for RNA secondary structure prediction~\cite{yin2025ernie}, we re-evaluated the same evaluation sequences across batch sizes 2, 4, 8, 16, 32, and 64 under dense padded inference.
Across batch sizes, batch context flipped 1.90--2.13\% of candidate pair decisions on average at a threshold of 0.5, despite only small perturbations in the continuous outputs, with mean logit MAE of 0.00382--0.00404 and mean probability MAE of $8.85\times10^{-4}$--$9.40\times10^{-4}$.
This effect was already present at batch size 2 and changed little as batch size increased, showing that reducing dense batch size does not eliminate the artifact.
Averaged across batch sizes, the flipped fraction was 1.83\% on Test, 2.24\% on GenA, 1.73\% on GenC, and 2.89\% on GenF (Fig.~\ref{fig:padding}c).
The largest relative disturbance occurred in GenF, indicating that the artifact is not restricted to the longest split.
Because the number of candidate pairs scales with sequence length, this corresponded to 72.0--79.3 flipped upper-triangular entries per sequence on average.
Dense padding therefore breaks batch-context invariance for variable-length RNA structure prediction even in small batches.

To remove this confound, we implemented a mask-aware NestedTensor execution path that excludes padded positions from the computational graph rather than masking them after dense computation.
We coupled this representation with a symmetry-aware triangular convolution that evaluates only the nonredundant half of the contact map while preserving symmetric output by construction (Fig.~\ref{fig:padding}d).
Padding burden and batch-context invariance were quantified on evaluation-set batch contexts from Test, GenA, GenC, and GenF, whereas memory and latency were benchmarked separately on synthetic batches spanning representative sequence lengths.
This design eliminates the mechanism underlying the batch-context effect by preventing padded positions from entering the computation, while simultaneously reducing wasted memory and arithmetic on padded and symmetric regions.

The computational gains were substantial even for the same U-Net architecture.
In the controlled systems benchmark, the padding-free implementation reduced the effective workload from 60.7 to 10.2 billion FLOPs, reduced allocated GPU memory from 4.61 to 0.69~GB, and reduced forward latency from 109.4 to 33.3~ms (Fig.~\ref{fig:padding}e).
This corresponds to a 3.3-fold inference speedup together with a 6.7-fold reduction in allocated memory.
During training, the same change reduced total step time from 229.5 to 115.0~ms and reduced peak training memory from 6.33 to 1.06~GB.
Padding-free computation therefore improves reproducibility, makes CHANRG-scale evaluation practical, and removes a hidden source of batch-context dependence from variable-length prediction (Fig.~\ref{fig:padding}c,d).

\section{Discussion}\label{sec:discussion}

CHANRG changes what counts as progress in RNA secondary-structure prediction.
Across a structure-aware, leakage-controlled benchmark, the models with the highest held-out base-pair accuracy did not generalize best to biologically distinct out-of-distribution regimes (Fig.~\ref{fig:main_results}a--c).
This divergence indicates that conventional held-out leaderboards can substantially overestimate robustness when structural redundancy and related-source leakage are not adequately controlled~\cite{qiu2023sequence,lasher2025bprna}.
By combining updated source coverage, structure-aware deduplication, genome-aware split construction, and hierarchical evaluation, CHANRG makes this gap directly measurable and provides a stronger basis for future method comparison.
These findings contrast with prior reports of strong generalization in RNA language models~\cite{penic2025rinalmo}, underscoring the importance of benchmark design.

The benchmark also reveals that high-capacity representation learning is not yet matched by equally transferable structural bias.
Foundation-model predictors achieved the strongest performance on the held-out Test split (Fig.~\ref{fig:main_results}a,c), consistent with strong within-distribution sequence-to-structure fitting.
However, their sharp decline on GenA, GenC, and GenF, together with the larger deterioration in topology $F_1$ and topology GED than in stem $F_1$, indicates that current foundation-model pipelines do not yet recover higher-order structure robustly outside familiar families (Fig.~\ref{fig:main_results}b,c; Fig.~\ref{fig:failure}a,d).
By contrast, structured decoders and the strongest direct neural baselines retained substantially more performance across OOD regimes in CHANRG (Fig.~\ref{fig:main_results}a,c; Fig.~\ref{fig:failure}a,d), suggesting that explicit structural constraints and task-aligned inductive bias remain important for transfer.
The scaling analysis supports the same interpretation, because increasing foundation-model capacity improved held-out accuracy much more than out-of-distribution robustness (Fig.~\ref{fig:length_scale}e).

CHANRG is therefore both a benchmark and a practical evaluation framework.
Its split design distinguishes held-out interpolation from transfer to architectural, clan-level, and genome-sparse family regimes (Fig.~\ref{fig:main}f), and its metric ladder separates local contact recovery from helix recovery, topology recovery, and higher-order structural damage (Fig.~\ref{fig:main}g; Fig.~\ref{fig:failure}a).
These distinctions make it possible to choose predictors according to the intended use case rather than according to a single held-out leaderboard.
For targets drawn from familiar and well-represented families, foundation-model predictors currently provide the highest held-out local structural accuracy (Fig.~\ref{fig:main_results}a).
For structurally novel or family-distant targets, however, structured decoders and the strongest direct neural predictors provide more reliable topology and more stable benchmark performance across OOD regimes (Fig.~\ref{fig:main_results}a,c; Fig.~\ref{fig:failure}a,d).
The accompanying padding-free, symmetry-aware reference implementation further makes CHANRG-scale evaluation batch-invariant, efficient, and practical for future model development and comparison (Fig.~\ref{fig:padding}a--d).

CHANRG is designed as a stringent stress test of structural generalization, not as an exhaustive taxonomy of RNA novelty.
Its OOD splits are operationally defined and capture specific forms of architectural and evolutionary shift, and structure-aware deduplication is only as complete as the structural representation used to define it.
In particular, the current framework provides the strongest control over non-crossing secondary-structure topology, whereas pseudoknot-involved interactions remain difficult to recover in current comparative evaluations~\cite{justyna2023rna}.
The benchmark likewise evaluates transfer through curated annotation and structural novelty rather than through direct downstream biological utility.
Future progress will therefore require not only larger or more expressive models, but also stronger transferable structural priors, richer benchmark tasks, and tighter links between benchmark robustness and experimentally validated RNA function.

\backmatter

\pagebreak

\bibliography{sn-bibliography}

\end{document}